\documentclass[12pt]{article}
\usepackage{epsfig}
\usepackage{latexsym}
\usepackage{bm}
\def\a{\alpha}
\def\b{\beta}
\def\d{\delta}
\def\th{\theta}
\def\k{\kappa}

\newcommand{\lam}{\lambda}
\newcommand{\del}{\partial}

\newcommand{\half}{{1 \over 2}}
\newcommand{\st}{{\tilde t}}
\newcommand{\sbo}{{\tilde b}}
\newcommand{\sq}{{\tilde q}}
\newcommand{\absv}[1]{\left|#1\right|}

\newcommand{\gtsim}{\mathrel{\hbox{\raise0.2ex
\hbox{$>$}\kern-0.75em\raise-0.9ex\hbox{$\sim$}}}}
\newcommand{\ltsim}{\mathrel{\hbox{\raise0.2ex
\hbox{$<$}\kern-0.75em\raise-0.9ex\hbox{$\sim$}}}}
\newcommand{\lw}[1]{\smash{\lower2.0ex\hbox{#1}}}

\newcommand{\Tr}{{\rm Tr}}

\newcommand{\expecv}[1]{\left\langle #1 \right\rangle}
\newcommand{\kslash}{k\kern-0.5em\raise 0.14ex\hbox{/}}
\def\barm{{\bar m}}

\def\RR{{\mathcal R}}
\def\MM{{\mathcal M}}
\def\II{{\mathcal I}}
\def\Veff{{V_{\rm eff}}}
\def\Iprime{{I${}^\prime$}}
\newcommand{\PRD}[3]{Phys. Rev. {\bf D{#1}} (19{#2}) {#3}}
\newcommand{\PRLet}[3]{Phys. Rev. Lett. {\bf {#1}} (19{#2}) {#3}}
\newcommand{\NPB}[3]{Nucl. Phys. {\bf B{#1}} (19{#2}) {#3}}
\newcommand{\PLB}[3]{Phys. Lett. {\bf B{#1}} (19{#2}) {#3}}
\newcommand{\PrTP}[3]{Prog. Theor. Phys. {\bf {#1}} (19{#2}) {#3}}
\makeatletter

\@addtoreset{equation}{section}
\makeatother
\begin{document}
\begin{titlepage}
\begin{flushright}
KYUSHU-HET 77\\
January~8,~2005
\end{flushright}
\vspace{50pt}
{\Large\bf
\begin{center}
Phase Transitions in the NMSSM
\end{center}}
\vskip2.0cm
\begin{center}
{\bf Koichi~Funakubo$^{a,}$\footnote{e-mail: funakubo@cc.saga-u.ac.jp},
Shuichiro Tao$^{b,}$\footnote{e-mail: tao@higgs.phys.kyushu-u.ac.jp}
and Fumihiko~Toyoda$^{c,}$\footnote{e-mail: ftoyoda@fuk.kindai.ac.jp}}
\end{center}
\vskip 0.8 cm
\begin{center}
{\it $^{a)}$Department of Physics, Saga University, Saga 840-8502 Japan}
\vskip 0.2 cm
{\it $^{b)}$Department of Physics, Kyushu University, Fukuoka 812-8581 Japan}
\vskip 0.2 cm
{\it $^{c)}$School of Humanity-Oriented Science and Engineering,
Kinki University, Iizuka 820-8555 Japan}
\end{center}
\vskip 1.5 cm
\centerline{\bf Abstract}
\vskip 0.2 cm
\baselineskip=15pt
We study phase transitions in the Next-to-Minimal Supersymmetric 
Standard Model (NMSSM) with the weak scale vacuum expectation values of the singlet scalar,
constrained by Higgs spectrum and vacuum stability.
We find four different types of phase transitions, three of which have two-stage nature.
In particular, one of the two-stage transitions admits strongly first order electroweak
phase transition, even with heavy squarks.
We introduce a tree-level explicit CP violation in the Higgs sector, which does not
affect the neutron electric dipole moment. In contrast to the MSSM with the CP violation
in the squark sector, a strongly first order phase transition is not so weakened 
by this CP violation.
\end{titlepage}
\baselineskip=16pt
\setcounter{page}{2}
\setcounter{footnote}{0}
%
%
\section{Introduction}
The origin of matter has been a long standing problem in astrophysics and particle physics.
Several attempts have been made to explain how the matter-antimatter asymmetry was
generated starting from the symmetric universe, in the early stage of the universe
before the nucleosynthesis.
Among such attempts, electroweak baryogenesis is a scenario which is intimately related to
physics at our reach\cite{reviewEWB}.
The scenario of electroweak baryogenesis, however, requires extension of the minimal standard
model, since the minimal standard model with the Higgs boson heavier than the present bound
does not give rise to strongly first-order phase transition, which is required to realize nonequilibrium
state, and CP violation in the CKM matrix is too small to generate sufficient baryon number.
Hence one needs extra sources of CP violation and light bosons, which strengthen 
the electroweak phase transition (EWPT).
The minimal supersymmetric standard model (MSSM) is one of such extensions of the standard model,
which can provide strongly first-order EWPT and contains many sources of CP violation.
The first-order EWPT is expected in the MSSM with a light top squark ($m_{\st_1}\le m_t$), 
which enhances $v^3$-behavior of the effective potential, where $v$ represents the expectation
value of the Higgs fields. The behavior is also enhanced for smaller $\tan\b$, that is, larger top Yukawa
coupling. The present bound on the lightest Higgs boson restricts the acceptable parameter space
of the model, which incorporates the strongly first-order EWPT.
Further, the first-order EWPT is weakened, when the stop sector CP violation, 
characterized by ${\rm Im}(\mu A_t e^{i\theta})$, is large, where $\theta$ is the relative phase of
the two Higgs doublets\cite{FunakuboTao}.
Thus, we are left with rather restricted range of the parameter region available for the baryogenesis
in the MSSM.\par
Another extension of the standard model is the next-to-minimal supersymmetric standard model
(NMSSM), which contains an extra singlet chiral superfield. It shares the advantages of the MSSM and
solves the $\mu$-problem inherent in the MSSM.
The NMSSM is reduced to the MSSM in the limit where the singlet decouples leaving
a finite $\mu$-term. For such peculiar parameters, we expect the same behavior of the EWPT
as that in the MSSM, which has been extensively studied.
 Here we shall focus on the parameter space far from the MSSM, that is,
the case where the vacuum expectation value of the singlet scalar has the magnitude of
the weak scale. One naturally expects such parameters, when all the soft masses of the scalars are
generated at the same scale where the supersymmetry is broken.
For such parameters, two of the authors studied Higgs spectrum with and without CP violation.
In \cite{FunakuboTao},   two distinct allowed parameter regions are chosen after imposing
conditions for the model to yield the sensible eletroweak-breaking vacuum.
The lightest Higgs boson for one type of the allowed parameter sets has moderate coupling to
$Z$ boson, while its mass is heavier than $114\mbox{GeV}$.
The other type of the parameter sets contains the Higgs bosons lighter than the bound,
but their couplings to the $Z$ boson are too small to be produced in the 
${\rm e}^-{\rm e}^+$ colliders. For such parameters, only the Higgs boson heavier than the bound
has the chance to be observed in future collider experiments.\par
Our purpose is to study finite-temperature phase transitions in the NMSSM with and without CP 
violation in the Higgs sector.
We find that the transitions in the case with light Higgs bosons are very different from that in the others.
This paper is organized as follows.
In Section~2, we introduce the model and define the parameters in the Higgs potential in a manner
independent of the phase conventions. The effective potential including at the one-loop level
is given in Section~3. Based on the symmetry of the effective potential, we discuss the possible
phases and transitions among them in Section~4. In Section~5, we present the numerical
results on the phase transitions with and without the explicit CP violation in the Higgs sector.
Section~6 is devoted to discussions on our results and their implication on the baryogenesis.
We summarize some formulas to calculate the effective potential in Appendix.
\section{The model}
The NMSSM is an extension of the MSSM, which contains one singlet superfield $N$.
The superpotential of the model is
\begin{equation}
 W = 
 \epsilon_{ij}\left( y_bH_d^iQ^jB - y_tH_u^iQ^jT + y_lH_d^iL^jE - \lambda N H_d^i H_u^j\right)
 -{\k\over3}N^3,
\end{equation}
which contains no dimensional parameter. The $\mu$-parameter in the MSSM is
induced when the scalar component of the singlet $N$ acquires nonzero vacuum
expectation value, $\mu=\lambda\expecv{N}$.
We shall not specify how the supersymmetry is broken, so that we introduce
generic type of soft supersymmetry-breaking terms.
The dimensional parameters such as the scalar masses, $A$-terms and gaugino masses,
are considered as inputs, which are constrained, for example, by vacuum stability conditions.
The soft supersymmetry-breaking terms are composed of those in the MSSM with $\mu=0$
and the terms including the singlet fields:
\begin{equation}
 {\mathcal L}_{\rm soft} = {\mathcal L}_{\rm soft}^{{\rm MSSM},\mu=0}
 -m_N^2 n^*n + \left[
 \lambda A_\lambda nH_dH_u + {\k\over3} A_\k n^3 + m'_N n^2 + \mbox{h.c.} \right]. \label{eq:soft-br-term}
\end{equation}
Among these terms the $n^2$-term breaks the global $\mbox{Z}_3$ symmetry, which causes
the domain wall problem upon broken spontaneously.
The $n^2$-term is not generated in the simple supergravity model, so we shall not include it in 
the following and consider that the discrete symmetry is explicitly broken by some higher dimensional
operator at very early era before the EWPT.\par
We parameterize the expectation values of the Higgs scalars as
\begin{equation}
  \expecv{\Phi_d} = \pmatrix{ {1\over{\sqrt2}}v_d\cr 0},\qquad
 \expecv{\Phi_u} = e^{i\theta}\pmatrix{0\cr {1\over{\sqrt2}}v_u},\qquad
 \expecv{n} = {1\over{\sqrt2}}e^{i\varphi}v_n,              \label{eq:def-VEVs}
\end{equation}
where $\Phi_d$, $\Phi_u$ and $n$ are the scalar components of $H_d$, $H_u$ and $N$, respectively.\par
The tree-level Higgs potential is written as a function of these expectation values:
\begin{eqnarray}
 V_0 &=&
 \half m_1^2 v_d^2 + \half m_2^2 v_u^2 + \half m_N^2 v_n^2 
 -\left[ {{\lambda A_\lambda}\over{2\sqrt2}}e^{i(\th+\varphi)}v_dv_uv_n +
          {{\k A_\k}\over{6\sqrt2}}e^{3i\varphi}v_n^3 + \mbox{h.c.} \right]    \nonumber\\
&&
 +{{g_2^2+g_1^2}\over{32}}(v_d^2-v_u^2)^2 +{{\absv{\lambda}^2}\over4}(v_d^2+v_u^2)v_n^2
 +\absv{ {\lambda\over2}e^{i\th}v_dv_u + {\k\over2}e^{2i\varphi}v_n^2 }^2,     \label{eq:V0}
\end{eqnarray}
where the terms in the first line consist of the soft-supersymmetry breaking terms, while
those in the second line come from the $D$- and $F$-terms.
Here we have four complex parameters $\lambda$, $\kappa$, $A_\lambda$ and $A_\k$,
and two phases $\theta_0$ and $\varphi_0$, which are the values of $\theta$ and
$\varphi$ evaluated at zero-temperature vacuum.
Some of these are redundant, so that we introduce the following notations in order to
present the results in a manner independent of phase conventions,
\begin{eqnarray}
 R_\lam \!\!\!&=&\!\!\! {1\over{\sqrt2}}{\rm Re}\left(\lam A_\lam e^{i(\th_0+\varphi_0)}\right), \qquad
 I_\lam = {1\over{\sqrt2}}{\rm Im}\left(\lam A_\lam e^{i(\th_0+\varphi_0)}\right),      \label{eq:def-RI-lambda}\\
 R_\k \!\!\!&=&\!\!\! {1\over{\sqrt2}}{\rm Re}\left(\k A_\k e^{3i\varphi_0}\right), \qquad\quad
 I_\k = {1\over{\sqrt2}}{\rm Im}\left(\k A_\k e^{3i\varphi_0}\right),         \label{eq:def-RI-kappa}\\
 \RR \!\!\!&=&\!\!\! {\rm Re}\left(\lam\k^* e^{i(\th_0-2\varphi_0)}\right),    \qquad\quad
 \II = {\rm Im}\left(\lam\k^* e^{i(\th_0-2\varphi_0)}\right).               \label{eq:def-RR-II}
\end{eqnarray}
As shown in \cite{FunakuboTao}, these parameters are constrained by the tadpole conditions,
which require the first derivatives of the potential to vanish at the prescribed vacuum.
In particular, CP-violation is characterized only by $\II$, since the conditions lead to
\begin{equation}
 I_\lam = {1\over2}\II v_{0n},\qquad
 I_\k = -{3\over2}\II {{v_{0d}v_{0u}}\over{v_{0n}}}=-{3\over2}\II {{v_0^2}\over{v_{0n}}}\sin\b_0\cos\b_0,
\end{equation}
where the subscripts $0$ denote the values evaluated at the vacuum and 
$\tan\b_0=v_{0u}/v_{0d}$.
When all the parameters are real, these equations are reduced to a trivial one.
If some of them are complex, we must arrange the parameters to respect this condition.
In practice, we give the absolute values of the parameters as inputs, then choose their
phases so as to satisfy this condition. 
As for the CP-even parameters, $R_\lam$, $R_\k$ and $\RR$ are related to the soft masses.
In favor of these parameters, one can eliminate $m_1^2$, $m_2^2$ and $m_N^2$ in terms of
$v_0$, $v_{0n}$ and $\tan\b_0$ by use of the tadpole conditions.\par
The MSSM limit, for which the singlet scalar decouples by taking $v_n\rightarrow\infty$ with
$\lambda v_n$ and $\kappa v_n$ fixed, the model is reduced to the MSSM, in the sense that
not only the spectrum but also the behavior of the EWPT are the same as those in the MSSM.
In the following, we focus on the case of weak scale $v_{0n}$, for which we expect new features
in the Higgs spectrum and the EWPT.
Then we obtain some new constraints on the parameters in the model.
In the MSSM, the vacuum parameterized by $v_0$ and $\tan\b_0$ is the global minimum of the
potential, as long as the tadpole conditions are satisfied.
On the contrary, in the NMSSM, we must require that the prescribed vacuum is the global minimum
of the potential, and masses squared of the other scalars, such as the charged scalars, evaluated there
are positive.
We also impose the condition that the lightest neutral Higgs mass is heavier than $114\mbox{GeV}$ when
its coupling to the $Z$ boson is not small.
These requirements select some allowed parameter sets from the vast parameter space.
The allowed parameter sets are classified into two kinds. The one contains a light Higgs
boson whose mass is smaller than $114\mbox{GeV}$ and the coupling to $Z$ boson is very small.
The other consists of the Higgs bosons, all of which are heavier than the bound.
We refer to a scenario with the former parameter sets as `light Higgs scenario'.
We found in \cite{FunakuboTao} that the allowed parameter region of the light Higgs type
disappears for $\tan\b_0\gtsim 10$. As we shall see later, the light Higgs scenario is realized
for small $\absv\k$ and always contain two light bosons.\par
In contrast to the MSSM, there is the tree-level CP violation in the Higgs sector, characterized by
the parameter $\II$. The CP violation in the Higgs sector of the MSSM is induced by
the loop corrections of the squarks and others. For example, ${\rm Im}\left(\mu A_q\right)$,
where $A_q$ is the squark $A$-term, generates the scalar-pseudoscalar mixing elements in
the neutral Higgs mass matrix. This CP violation also affects the neutron EDM, so that the
squark mass and/or the relative phase of $\mu$ and $A_q$ are constrained not to exceed
the upper bound on the EDM. The effect of this CP violation on the Higgs spectrum and 
gauge couplings have been studied by Carena, et al.\cite{carena}. They found that sufficiently
large CP violation mixes the scalar and pseudoscalar, so that the lightest Higgs boson has
very small gauge coupling to $Z$ boson to escape from the present mass bound.
Although one expects strongly first-order EWPT with such a light boson, 
we found that the CP violation weakens the EWPT, which was strongly first order because of
the light stop in the CP-conserving case\cite{FTT-1}.\par
On the other hand, the CP violation in the Higgs sector of the NMSSM can escape
from the bound due to the EDM\cite{FunakuboTao},
while it can generate the CP-violating bubble wall created at the first-order EWPT.
We shall see the effect of the tree-level CP violation on the EWPT below.
\section{Effective potential}
In addition to the tree-level potential (\ref{eq:V0}), we include the one-loop corrections coming
from the loops of the third generation of quarks and squarks, the gauge bosons and the singlet fermion.
The mass-squared matrix of the neutral Higgs bosons and the mass of the charged Higgs boson
are defined as the second derivatives of the effective potential evaluated at the vacuum.
The explicit forms of these derivaties are given in \cite{FunakuboTao}.\par
The finte-temperature effective potential is given by
\begin{equation}
 \Veff(\bm{v};T) = V_0(\bm{v}) + \Delta V(\bm{v};T),            \label{eq:def-Veff}
\end{equation}
where $V_0$ is the tree-level Higgs potential (\ref{eq:V0}). The one-loop correction is
\begin{eqnarray}
 \Delta V(\bm{v};T) \!\!\!&=&\!\!\!
 3\left[ F_0(\barm_Z^2) + {{T^4}\over{2\pi^2}}I_B\!\!\left({{\barm_Z}\over T}\right)
        +2F_0(\barm_W^2) + 2\cdot{{T^4}\over{2\pi^2}}I_B\!\!\left({{\barm_W}\over T}\right) \right]   \nonumber\\
&&
 - 2\left[ F_0(\barm_{\psi_N}^2) + {{T^4}\over{2\pi^2}}I_F\!\!\left({{\barm_{\psi_N}}\over T}\right) \right]      \nonumber\\
&&
 +N_C\sum_{q=t,b}\left\{ 
   2\sum_{j=1,2}\left[ F_0(\barm_{\sq_j}^2) + {{T^4}\over{2\pi^2}}I_B\!\!\left({{\barm_{\sq_j}}\over T}\right)\right]
   -4\left[ F_0(\barm_q^2) + {{T^4}\over{2\pi^2}}I_F\!\!\left({{\barm_q}\over T}\right) \right] \right\},
       \nonumber\\
                                                  \label{eq:Delta-V}
\end{eqnarray}
where $\barm$ denotes the field-dependent masses, given in Appendix~A and
\begin{eqnarray}
 F_0(m^2) &=& {1\over{64\pi^2}}\left(m^2\right)^2\left(\log{{m^2}\over{M^2}} - {3\over2}\right),      \\
 I_B(a) &=& \int_0^\infty dx\, x^2\log\left( 1 - e^{-\sqrt{x^2+a^2}} \right),     \\
 I_F(a) &=& \int_0^\infty dx\, x^2\log\left( 1 + e^{-\sqrt{x^2+a^2}} \right).      \label{eq:def-F-B-F}
\end{eqnarray}
Here we adopt the $\overline{{\rm DR}}$ scheme with the renormalization scale $M$, which
we determined in such a way that the one-loop correction to the potential vanishes
at the vacuum.
The order parameters $\bm{v}=(v_1, v_2, v_3, v_4, v_5)$ are related to the expectation values
of the Higgs fields as follows,
\begin{equation}
 \left(v_1,v_2,v_3,v_4,v_5\right) =
 \left( v_d,\, v_u\cos\Delta\th,\, v_u\sin\Delta\th,\, v_n\cos\Delta\varphi,\, v_n\sin\Delta\varphi \right),
                                 \label{eq:def-vector-v}
\end{equation}
where $\Delta\th = \th-\th_0$ and $\Delta\varphi=\varphi-\varphi_0$.
Then the tree-level potential is expressed as
 \begin{eqnarray}
 V_0(\bm{v}) \!\!\!&=&\!\!\!
 \half m_1^2 v_1^2 + \half m_2^2(v_2^2+v_3^2) + \half m_N^2(v_4^2+v_5^2)     \nonumber\\
 &&
 -\left[R_\lam(v_2v_4-v_3v_5)- I_\lam(v_3v_4+v_2v_5)\right] v_1
 -{1\over3}\left[ R_\k(v_4^2 -3v_5^2)v_4 - I_\k(3v_4^2 - v_5^2)v_5 \right]          \nonumber\\
 &&
 +{{g_2^2+g_1^2}\over{32}}(v_1^2-v_2^2-v_3^2)^2
 +{{\absv\lam^2}\over4}\left[ (v_1^2+v_2^2+v_3^2)(v_4^2+v_5^2)+v_1^2(v_2^2+v_3^2)\right]  \nonumber\\
 &&
 +{{\absv\k^2}\over4}(v_4^2+v_5^2)^2 +
 {{v_1}\over2}\left[ \RR\left( v_2(v_4^2-v_5^2) + 2v_3v_4v_5 \right) - \II\left(v_3(v_4^2-v_5^2) -2v_2v_4v_5\right)
             \right].    \nonumber\\
 &&        \label{eq:V0-vector-v}
\end{eqnarray}
The tadpole conditions recieve the one-loop corrections, which modify the expression for the soft masses and the CP-violating parameter $I_\lambda$.
The results are summarized in Appendix~A.\par
When we examine the consistency of the parameters and study the EWPT, we search for
the minimum in the five-dimensional space of the order parameters $\bm{v}$.
Noting that the tree-level potential $V_0$ and all the field-dependent masses are invariant under
\begin{equation}
  (v_1,v_2,v_3,v_4,v_5)\; \mapsto\; (-v_1,-v_2,-v_3,v_4,v_5),         \label{eq:discrete-sym-Veff}
\end{equation}
the effective potential also has this discrete symmetry.
Hence, it is sufficient to search for the minimum in the space with $v_u\ge0$.
This symmetry implies that the first derivatives of the effective potential with respect to $v_1$, $v_2$
and $v_3$ at the origin vanish at any temperature.\par
The phase transitions are studied by searching for minima of the effective potential at each temperature.
Among the phase transitions, the EWPT is defined as the transition at which $v(T)\equiv\sqrt{v_1^2(T)+v_2^2(T)+v_3^2(T)}$
vanishes when temperature is raised, irrespective of $v_n(T)\equiv\sqrt{v_4^2(T)+v_5^2(T)}$,
where $\bm{v}(T)$ denotes the absolute minimum of the effective potential at temperature $T$.
For successful electroweak baryogenesis, the EWPT must be strongly first order, in order for the 
generated baryon number not to be washed out by the sphaleron process.
In the MSSM, this requires a light stop, whose mass is less than the top quark mass.
Then, the contribution of the light stop to the effective potential, proportional to $T^4 I_B(\bar m_{\st_1}/T)$,
produces an effective $T v^3$-term with a negative coefficient. 
Althought such terms also come from the gauge boson loops, the stop contribution is much larger 
because of the large Yukawa coupling and the color degrees of freedom\cite{lightstop}.
Such a light stop is realized when the singlet stop soft mass almost vanishes, while 
the doublet soft mass cannot be taken so small at the same time to avoid too large deviation
of the $\rho$-parameter from unity. Hence, a successful baryogenesis in the MSSM requires
a specific breaking of supersymmetry.\par
In the NMSSM, the scalar trilinear term in the tree-level potential is expected to yield strongly first order 
EWPT\cite{pietroni}. 
Indeed, the EWPT was studied for a wide range of parameter space and was found to 
have more chances to be first order for smaller Higgs mass\cite{davies}.
The naive argument of \cite{pietroni} is as follows:
If the EWPT proceeds along the straight line in the order parameter space, connecting the origin
and the minimum of the effective potential corresponding to the broken phase, 
one can parameterize the order parameters as
\begin{eqnarray}
 v_d &=& v\cos\b(T) = y\cos\a(T)\cos\b(T),          \nonumber\\
 v_u &=& v\sin\b(T) = y\cos\a(T)\sin\b(T),            \label{eq:straight-paramet-vs}\\
 v_n &=& y\sin\a(T),     \nonumber
\end{eqnarray}
with (almost) constant $\a(T)$ and $\b(T)$ at each temperature $T$.
Then the tree-level potential is written as
\begin{eqnarray}
 V_0 &=&
 \half\left( (m_1^2\cos^2\b+m_2^2\sin^2\b)\cos^2\a+m_N^2\sin^2\a\right) y^2    \nonumber\\
 &&\qquad
 - \left( R_\lambda\cos^2\a\sin\a \cos\b\sin\b +{1\over3}R_\kappa\sin^3\a\right) y^3 +\cdots.
\end{eqnarray}
For appropriate parameters, the coefficient of the $y^3$-term is negative, which makes the phase
transition along $y$-direction strongly first order.
In the MSSM, the EWPT proceeds along an almost constant-$\beta$ line\cite{mssmEWPT}.
Since there is no specific symmetry among the doublets and the singlet,
the validity of the parametrization (\ref{eq:straight-paramet-vs}) with a constant $\a$
is not obvious. 
As we shall see below, the phase transitions in the NMSSM are classified into
several types, only one of which admits the parameterization of the order parameters in 
(\ref{eq:straight-paramet-vs}).
\section{Possible phases and transitions}
Before presenting the numerical results, we discuss the possible phases and transitions among them.
A phase at each temperature is characterized by the location of the minimum of the effective potential
$\bm{v}(T)$.
Among the components of $\bm{v}$, $v$ and $v_n$ are related to the symmetries of the model.
Obviously $v$ is the order parameter of the gauge symmetry, while $v_n$ is the order parameter
of a global $U(1)$ symmetry. This is because in the subspace of $v_n=0$ ($v_4=v_5=0$), 
the effective potential is invariant under the global $U(1)$ transformation $v_2+iv_3\mapsto e^{i\alpha}(v_2+iv_3)$.
We denote the phases of different symmetries as listed in Table~\ref{tab:phases}.
\begin{table}[h!]
\begin{center}
\begin{tabular}{c||c|c}
\hline
 phase & order parameters & symmetries \\
\hline
 EW  & $v\not=0$, $v_n\not=0$ & fully broken    \\
 I, \Iprime      & $v=0$, $v_n\not=0$ &  local $SU(2)_L\times U(1)_Y$ \\
 II     & $v\not=0$, $v_n=0$ & global $U(1)$   \\
 SYM & $v=v_n =0$  & $SU(2)_L\times U(1)_Y$, global $U(1)$  \\
\hline
\end{tabular}
\end{center}
\caption{Classification of the phases in the NMSSM.}
\label{tab:phases}
\end{table}
For any sensible parameters, the model is in the phase-EW at much lower temperatures than
the electroweak scale.
At sufficiently high temperatures, where all the symmetries are restored, it is in the phase-SYM.
In the limit where the singlet fields decouple by $v_n\rightarrow\infty$, the phase-EW at low
temperatures transits to the phase-I with almost constant $v_n(T)$, resulting in the MSSM-like EWPT. 
In the case of $v_{0n}=O(100)\mbox{GeV}$, all the phases can appear at temperature of the electroweak
scale so that various patterns of phase transitions will be found.\par
The existence of the phases-I and II is a novel feature of the NMSSM.
In particular, in the subspace of $v=0$, the potential can develop a nontrivial minimum for some
parameters. To see this fact, we consider the tree-level potential for $v_1=v_2=v_3=0$,
\begin{equation}
 \hat V_0(v_n) = V_0(0,0,0, v_n\cos\Delta\varphi, v_n\sin\Delta\varphi)
  = {1\over2}m_N^2 v_n^2 -{1\over3}\hat R_\k v_n^3 + {{\absv{\k}^2}\over4}v_n^4,  \label{eq:def-V0-hat}
\end{equation}
where $\hat R_\k = {1\over{\sqrt2}}\absv{\k A_\k}\cos[{\rm Arg}(\k A_\k)+3\Delta\varphi]$.
When $\hat R_\k^2> 4\absv{\k}^2m_N^2$, $\hat V_0(v_n)$ has a local minimum at
$v_n=\alpha_+= (\hat R_\k+\sqrt{\hat R_\k^2-4\absv{\k}^2m_N^2})/(2\absv{\k}^2)$, where the potential is
\begin{equation}
 \hat V_0(\alpha_+) = 
 {{\a_+^2}\over4}\left( 
  m_N^2 - {{\hat R_\k^2+\hat R_\k\sqrt{\hat R_\k^2-4\absv{\k}^2m_N^2}}\over{12\absv{\k}^2}} \right).
                         \label{eq:hatV0-alpha}
\end{equation}
For a very small $\absv\k$ with moderate $\absv{A_\k}$ and $m_N^2$, 
$\a_+$ is so large that $\hat V_0(\a_+)$ becomes smaller than the potential at the electroweak vacuum, when the right-hand side of (\ref{eq:hatV0-alpha}) is negative.
That is, for such a parameter set, the phase at zero temperature is the phase-I.
Although we must exclude such a parameter set, some parameters can admit the phase-I of this kind
as an intermediate phase at finite temperature.
We shall refer to the phase realized for a small $\absv\k$ as phase-\Iprime, in order to distinguish it
from the phase-I which occurs as the intermediate phase in the MSSM-like phase transition.
Both the phases will appear at finite temperature with the same symmetry property.
However, the Higgs spectrum for the parameters which admits the phase-\Iprime{}  is different from
that in the MSSM-like parameters, in that the former contains a light scalar.
In the limit $\k=0$, the model is invariant under a phase transformation of $N$ and $H_u$,
which is spontaneously broken by nonzero $v_{0n}$. This symmetry is explicitly broken by small $\absv\k$,
which results in a light Higgs boson.
It should be noted that this argument is not so altered by the radiative corrections.
If we set $v_1=v_2=v_3=0$ in the effective potential, the correction dependent on $v_n$
arises only from the loop of the singlet fermion, whose contribution is negligible for $\absv\k\ll 1$.
This also implies that the value of the effective potential in the phase-\Iprime{}  is almost independent
on temperature.\par
As temperature is lowered, the phase-SYM at high temperature transits to the phase-EW via various phases.
We enumerate the four types of transitions which we encounter in the numerical analysis:\\[4pt]
\hspace{30mm}
\begin{minipage}[t]{120mm}
\begin{tabbing}
 \hspace*{50mm} \= \hspace*{50mm} \kill
 A: SYM $\rightarrow$ I $\Rightarrow$ EW  \> B: SYM $\rightarrow$ \Iprime{}  $\Rightarrow$ EW  \\[4pt]
 C: SYM $\Rightarrow$ II $\rightarrow$ EW \> D: SYM $\Rightarrow$ EW
\end{tabbing}
\end{minipage}\\[4pt] \noindent
Here the double arrow indicates the transition at which the electroweak gauge symmetry is broken.
The second transition of type~A is the MSSM-like EWPT, which proceeds with almost constant $v_n(T)$.
The transition of type~D was found to be strongly first order transition in the case of light Higgs boson\cite{pietroni}.
We shall see below that such a transition for the parameters consistent with the lower bound on the Higgs mass
occurs in the presence of a light stop.
The other two-stage transitions are novel in the NMSSM.
In principle, there could be a three-stage transition, but it will be realized for very restricted parameters.

\section{Numerical results}
The behavior of the phase transitions is studied by numerically searching for the minimum of the effective
potential (\ref{eq:def-Veff}). We do not use the high-temperature expansion to evaluate the integrals
(\ref{eq:def-F-B-F}), since there are particles of weak scale mass.
The method we employed is the same as that in \cite{FTT-1}.
We perform the minimum search of the effective potential for various parameter sets.
Among the parameters, the common ones are those known from experiments such as 
$v_0=246\mbox{GeV}$, $m_W=80.3\mbox{GeV}$, $m_Z=91.2\mbox{GeV}$, $m_t=175\mbox{GeV}$, 
$m_b=4.2\mbox{GeV}$, and we adopt the $A$-parameters in the squark sector $A_t=A_b=20\mbox{GeV}$,
which are taken not so large that the squark masses-squared do not become negative.
As for the squark soft masses, we take the following three cases: in the heavy squark case 
the doublet soft mass is taken to be $m_\sq=1000\mbox{GeV}$ and the singlet ones are
$m_{\st_R}=m_{\sbo_R}=800\mbox{GeV}$. In the light squark case-I, $m_\sq=1000\mbox{GeV}$ and
$m_{\st_R}=m_{\sbo_R}=10\mbox{GeV}$, and in the light squark case-II,
$m_\sq=500\mbox{GeV}$ and the same singlet soft mass as the case-I.
Now the remaining parameters are those characterizing the vacuum $\tan\b_0$, $\theta_0$,
$\varphi_0$ and $v_{0n}$, and
the complex ones, $\lambda$, $\kappa$, $A_\lambda$, $A_\k$.
The soft masses in the Higgs potential can be expressed in terms of the other parameters by use
of the tadpole conditions, as shown in Appendix~A.
As discussed in Section~3, the independent parameters in the Higgs sector
are $\absv{\lambda}$, $\absv{\k}$, $R_\lambda$, $R_\kappa$, $\RR$ and $\II$.
We use the charged Higgs mass $m_{H^\pm}$, instead of $R_\lambda$, through the relation
\begin{equation}
 m_{H^\pm}^2 = m_W^2-{1\over2}\absv{\lambda}^2v_0^2+
 {1\over{\sin\b_0\cos\b_0}}\left[ \left( R_\lambda-{1\over2}\RR v_{0n}\right)v_{0n}
 + \expecv{ {{\del^2 \Delta V(\bm{v};0)}\over{\del\phi_d^+\del\phi_u^-}} } \right],
\end{equation}
where $\expecv{\cdots}$ represents the value evaluated at the vacuum.
The expression of the second derivative with respect to the charged Higgs field is rather
lengthy and almost the same as that in the MSSM, which is given in \cite{FTT-1}, except for 
$\mu$-parameter replaced with $\lambda v_{0n} e^{i\varphi_0}/\sqrt2$.\par
\begin{figure}[h!]
\centerline{\epsfig{file=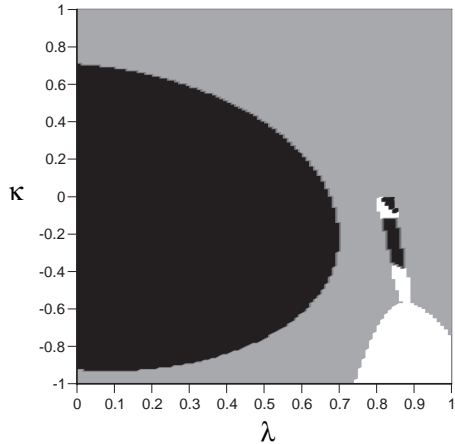, height=60mm}}
\caption{The allowed region in the $(\lambda,\k)$-plane for $v_{0n}=200\mbox{GeV}$, $\tan\b_0=5$, 
$A_\k=-100\mbox{GeV}$ and $m_{H^\pm}=600\mbox{GeV}$ in the heavy squark case.}
\label{fig:crit-5h-600}
\end{figure}

We thoroughly examined the parameter space of the model for $v_{0n}=100-1000\mbox{GeV}$
in the absence of CP violation\cite{FunakuboTao}.
The allowed parameter sets are defined to satisfy the two conditions; i) the Higgs bosons whose
coupling to the $Z$ boson is larger than a tenth of that in the standard model Higgs boson be
heavier than $114\mbox{GeV}$ (the spectrum condition), and ii) the prescribed vacuum be
the absolute minimum of the effective potential (the vacuum condition).
The second condition also requires that all the masses-squared of the scalars in the model be
positive, including the squarks and sleptons.
The allowed regions in the parameter space are roughly classified into two groups; one  conains a
light Higgs boson with small coupling to the $Z$ boson, and the other is similar to the MSSM, with 
the lightest Higgs boson heavier than the present bound.
The former exists only when $v_{0n}$ is $O(100)\mbox{GeV}$, and within the region, $\absv\k\ltsim 0.1$.
The latter region always exists for a large $\absv\k$ regime when $v_{0n}$ is small, and extends to
small $\absv\k$ regime as $v_{0n}$ grows to reach the MSSM limit.
In the following, we show the results on the phase transitions for several points within the allowed regions
for $v_{0n}=200\mbox{GeV}$, $\tan\b_0=5$ and $A_\k=-100\mbox{GeV}$, to illustrate the four types of the transitions
discussed in the previous section in the absence of CP violation.
For $m_{H^\pm}=600\mbox{GeV}$ in the heavy squark case, the allowed region in the $(\lambda,\k)$-plane
is shown as a white region in Fig.~\ref{fig:crit-5h-600}, where the grey region is excluded by the spectrum condition
and the black region is excluded by the vacuum condition\cite{FunakuboTao}.
Later, we introduce the explicit CP violation in the Higgs sector which does not induce the neutron EDM
and examine its effects on the phase transitions.
%
%
%
\subsection{CP-conserving case}
We take all the parameters to be real and set $\th_0=\varphi_0=0$.
For $v_{0n}=200\mbox{GeV}$, $\tan\b_0=5$ and $A_\k=-100\mbox{GeV}$, we can find all the four types of
phase transitions by choosing an appropriate set of $m_{H^\pm}$, $\lambda$, $\k$ and the squark soft masses.
The transition of type~D is observed only with the light squark cases in the light Higgs scenario.
The transition of type~A is found in rather broad region with the heavy Higgs boson, but a light stop is
necessary for the EWPT to be strongly first order just as in the case of the MSSM.
The transitions of type~B and C appear in the light Higgs scenario at $\absv{\k}\ll 1$.
The type-C transition also requires a light stop to be strongly first order.
Here we take $(\lambda,\k)=(0.9,-0.9)$ and $(0.82, -0.05)$ with $m_{H^\pm}=600\mbox{GeV}$
in the light squark-I case. Each corresponds to the transition of type~A and C, respectively.
An example of the type-B transition, we choose $(\lambda,\k)=(0.85,-0.1)$ with $m_{H^\pm}=600\mbox{GeV}$
in the heavy squark case.
For an example of the type-D transition, we adopt $(\lambda,\k)=(0.96,-0.02)$ and $m_{H^\pm}=700\mbox{GeV}$
in the case of the light squark-II.
We shall refer to each parameter set as A, C, B, and D, respectively.
The masses and $g_{HZZ}^2$ of the mass eigenstates are listed inTable~\ref{tab:higgs-mass-CP-cons} for each 
parameter set. Here $g_{HZZ}$ is the couplings to the $Z$ boson normalized by that of the standard model.
\newcommand{\lowent}[1]{\smash{\lower1.5ex\hbox{#1}}}
\begin{table}[h!]
\begin{center}
\begin{tabular}{c|c|ccccc}
\hline
       & & $H_1$ & $H_2$ & $H_3$ & $H_4$ & $H_5$    \\
\hline\hline
  \lowent{A} 
  & $m_{H_i}$(GeV) & $119.53$ & $203.59$ & $265.74$ & $617.24$ & $637.47$  \\
  & $g_{H_iZZ}^2$  & $0.9992$  & $5.926\times10^{-4}$ & $0$ & $0$ & $1.884\times10^{-4}$ \\
\hline
  \lowent{B} 
  & $m_{H_i}$(GeV) &$38.89$ & $75.31$ & $131.11$ & $625.61$ & $627.95$  \\
  & $g_{H_iZZ}^2$  & $6.213\times10^{-8}$  & $0$ & $0.9999$ & $6.816\times10^{-5}$ & $0$ \\
\hline
  \lowent{C} 
  & $m_{H_i}$(GeV) & $42.24$ & $63.49$ & $117.25$ & $625.09$ & $627.44$  \\
  & $g_{H_iZZ}^2$  & $0.00188$  & $0$ & $0.9980$ & $9.541\times10^{-5}$ & $0$ \\
\hline
  \lowent{D} 
  & $m_{H_i}$(GeV) & $41.88$ & $58.62.08$ & $115.15$ & $730.51$ & $734.58$  \\
  & $g_{H_iZZ}^2$  & $0$  & $1.015\times10^{-4}$ & $0.9997$ & $1.632\times10^{-4}$ & $0$ \\
\hline
\end{tabular}
\end{center}
\caption{The mass and $g_{HZZ}^2$ of the Higgs mass eigenstates for the four parameter sets.}
\label{tab:higgs-mass-CP-cons}
\end{table}
For these parameter sets,
the only one CP-even scalar has the coupling almost equal to unity, while the others have almost zero.
This is a general feature when the charged scalar is rather heavy, since the tree-level mass matrix of 
the CP-even scalar has an eigenvector near $(\cos\b_0, \sin\b_0,0)$ so that the corresponding mass
engenstate has $g_{HZZ}^2\simeq1$\cite{FunakuboTao}.
The mass eigenvalues are also be understood by recalling the properties of the tree-level mass
matrices as follows. The mass eigenvalues of the pseudoscalar always staisfy
$m_{A_1}^2 < \hat m^2 < m_{A_2}^2$, where 
$\hat m^2= m_{H^\pm}^2-m_W^2+\absv{\lambda}^2v_0^2/2$. When $\hat m^2$ is large,
the mass eigenvalues of the CP-even scalars satisfy
$m_{S_1}^2<m_{S_2}^2<\hat m^2<m_{S_3}^2$.
As discussed in the previous section, there is a light pseudoscalar $A_1$ for small 
$\absv\k$ and $v_{0n}\not=0$.
Then the relations $\Tr\MM_S^2\simeq \hat m^2(1+v_0^2\sin^2\b_0\cos^2\b_0/v_{0n}^2)+m_Z^2$
and $\Tr\MM_P^2\simeq \hat m^2(1+v_0^2\sin^2\b_0\cos^2\b_0/v_{0n}^2)$ imply that
$m_{A_2}\simeq m_{S_3}\simeq \hat m$, and that $m_{S_1}$ must be smaller than $m_Z$,
once we impose the condition $m_{S_2}>114\mbox{GeV}$.
Thus the spectrum of the neutral Higgs bosons listed in Table~\ref{tab:higgs-mass-CP-cons}
is generic for $\absv\k\simeq1$ (set~A) and $\absv\k\ll 1$ (B, C, D), when the charged scalar is heavy.
This should be contrasted to the situation in the MSSM, in which the pseudoscalar and
the heavier scalar decouple leaving the lighter scalar at the weak scale,
 in the limit of large $m_{H^\pm}$.\par
\begin{figure}[h!]
\centerline{\epsfig{file=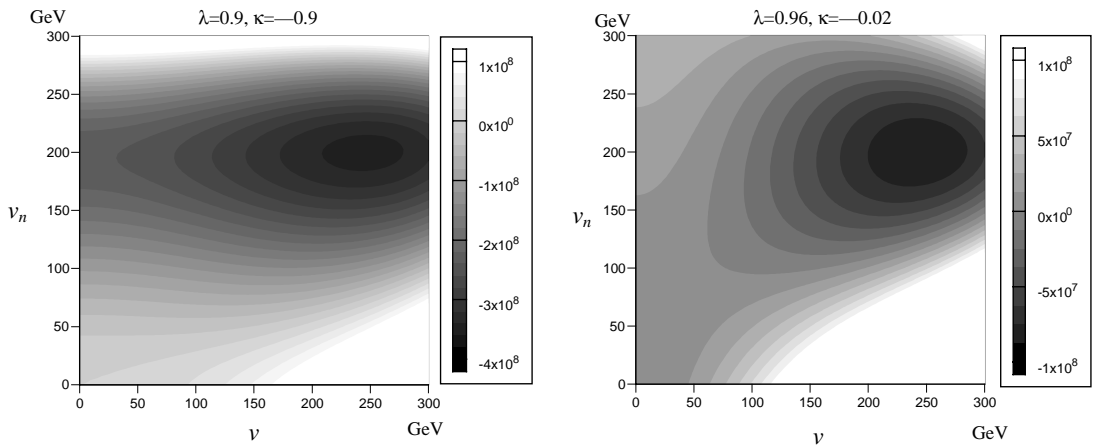, height=60mm}}
\caption{The contour plot of the reduced effective potential at $T=0$ for the parameter 
set~A (left-hand side) and D (right-hand side).}
\label{fig:red-Veff-AD}
\end{figure}
\begin{figure}[h!]
\centerline{\epsfig{file=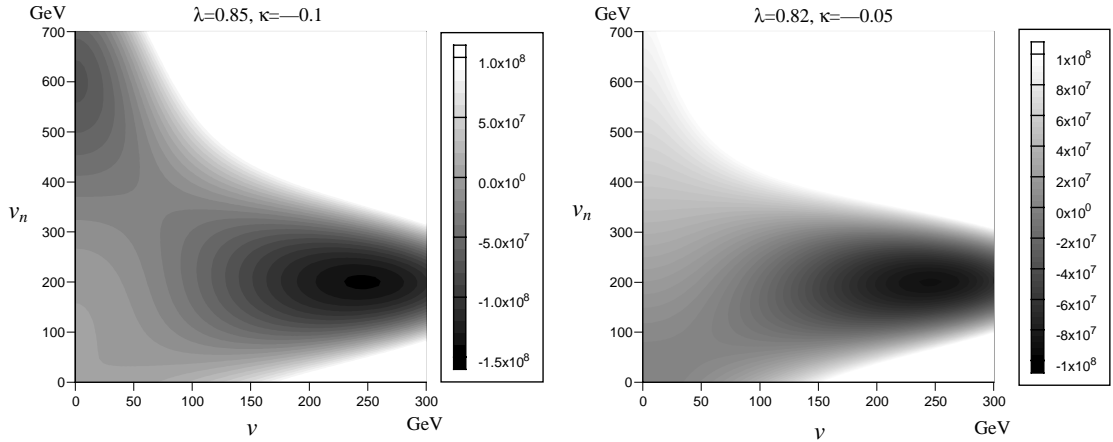, height=60mm}}
\caption{The contour plot of the reduced effective potential at $T=0$ for the parameter 
set~B (left-hand side) and C (right-hand side).}
\label{fig:red-Veff-BC}
\end{figure}

The phase transitions are studied by searching for the global minimum of the effective potential,
which is a function of $(v_1, v_2, v_4)=(v_d, v_u, v_n)$ in the CP-conserving case. 
Which type of transitions is realized is anticipated from the structure of the zero-temperature
potential. To show the potential schematically, we define the reduced effective potential as
a function of $(v,v_n)$, which is a section of the three-dimensional order parameter space
with a constant $\tan\b(T)$ at the minimum,
\begin{equation}
 \tilde V_{\rm eff}(v, v_n;T) = \Veff( v\cos\b(T), v\sin\b(T), 0, v_n, 0; T)-\Veff( 0, 0, 0, 0, 0; T).
\end{equation}
The contour plots of the reduced potential at zero temperature for the parameter sets~A--D are 
shown in Figs.~\ref{fig:red-Veff-AD} and \ref{fig:red-Veff-BC}.
The left-hand plot in Fig.~\ref{fig:red-Veff-AD} suggests that the phase transition proceeds
along almost constant $v_n$, while the right-hand plot seems to justify the parametrization
of (\ref{eq:straight-paramet-vs}) with a constant $\a(T)$.
The left-hand plot in Fig.~\ref{fig:red-Veff-BC} shows there is a local minimum along $v=0$.
As discussed in the previous section, the height of that minimum is almost indepenedent of
$T$, while that at the vacuum grows as temperarture is raised, so that the transition will be
of the type-B.
To show the global structures, we adopt a larger scale for the vertical axes in Fig.~\ref{fig:red-Veff-BC}
than those in Fig.~\ref{fig:red-Veff-AD}.
If we used the same scale for them, the contour plot for the set~C would looks like that for the set~D.\par
\begin{figure}[h!]
\centerline{\epsfig{file=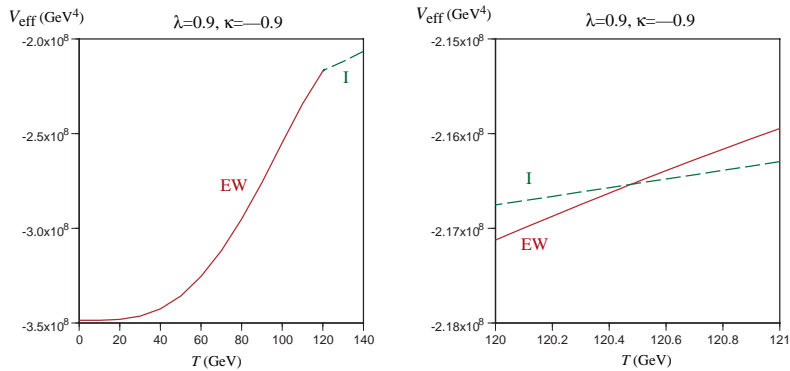, height=50mm}}
\caption{The effective potentials at the local minima corresponding to the phase-EW (solid curve)
and the phase-I (dashed curve) for the parameter set~A. The right-hand plot is the close view of the left-hand one
near the transition temperature.}
\label{fig:Veffs-A}
\end{figure}

To see how the transitions proceed, we show the temperature dependence of the local minima
of the effective potential corresponding to the phases, subtracted by the value at the origin.
In the numerical search for the minima, we found a few distinct convergent points 
at the same temperature, which are displayed in the plots as different curves.
For the set~A, the result is plotted in Fig.~\ref{fig:Veffs-A}, which shows that the phase-EW
changes to the phase-I at $T_C=120.47\mbox{GeV}$. There the order parameters change
from $(v,v_n)=(106.92\mbox{GeV}, 194.23\mbox{GeV})$ to $(0, 192.75\mbox{GeV})$.
This implies the EWPT is first order with $v_C/T_C=0.89$. Although we cannot determine the sphaleron
decoupling condition in the NMSSM, since we have not found the sphaleron solution in the model,
that will be similar to that in the MSSM for the set~A, whose  spectrum and phase transition are
MSSM-like. Then this parameter will not provide sufficiently strong first-order transition, but
the transition will become stronger for a lighter stop, as in the case of the MSSM.\par
\begin{figure}[h!]
\centerline{\epsfig{file=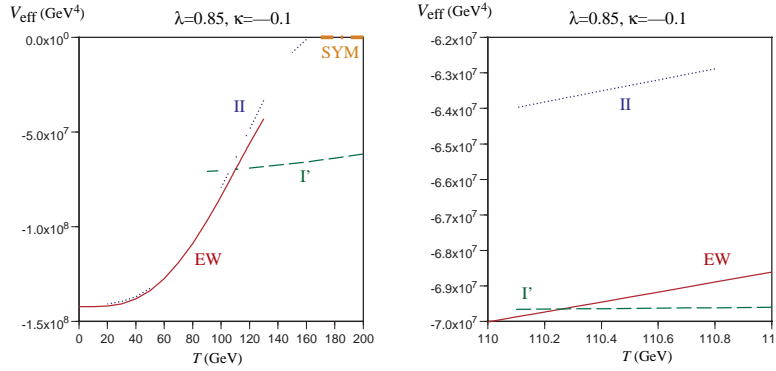, height=50mm}}
\caption{The effective potentials at the local minima corresponding to the phase-EW (solid curve),
the phase-\Iprime{}  (dashed curve), the phase-II (dotted curve) and the phase-SYM (thick dotted-dashed
curve) for the parameter set~B. The right-hand plot is the close view of the left-hand one near the transition temperature.}
\label{fig:Veffs-B}
\end{figure}

An example for the transition of type~B is shown in Fig.~\ref{fig:Veffs-B}.
At $T_C=110.26\mbox{GeV}$, the low-temperature phase-EW transits to the phase-\Iprime, 
which gives its place to the phase-SYM at $T>500\mbox{GeV}$.
There the order parameters changes from $(v,v_n)=(208.13\mbox{GeV}, 248.85\mbox{GeV})$ to
$(0, 599.93\mbox{GeV})$.
This is a very strongly first-order transition with respect to the electroweak order parameter.
As discussed in the previous section, the slopes of the curves corresponding to the phase-I and \Iprime{}
are smaller than the others. In particular, the potential at the phase-\Iprime{}  is amost independent of
temperature because of small $\absv\k$.\par
\begin{figure}[h!]
\centerline{\epsfig{file=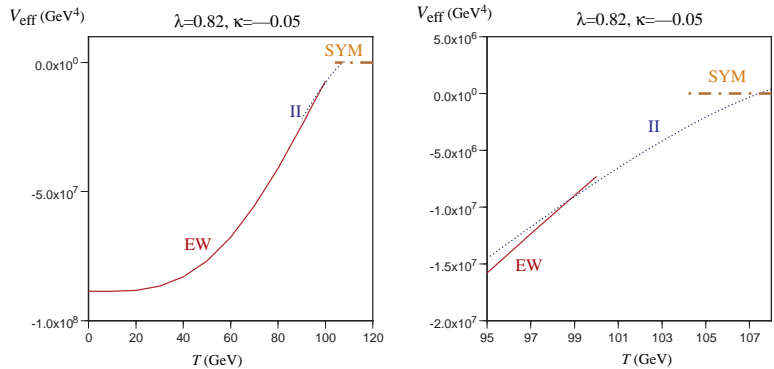, height=50mm}}
\caption{The effective potentials at the local minima corresponding to the phase-EW (solid curve),
the phase-II (dotted curve) and the phase-SYM (thick dotted-dashed curve) for the parameter set~C. 
The right-hand plot is the close view of the left-hand one near the transition temperature.}
\label{fig:Veffs-C}
\end{figure}

Fig.~\ref{fig:Veffs-C} illustrates the transition of type~C, in which the low-temperature phase-EW
changes to the phase-SYM through the intermediate phase-II.
At $T_N=98.76\mbox{GeV}$, the phase-EW changes to the phase-II, which is converted to the phase-SYM
at $T_C=107.44\mbox{GeV}$. The order parameters changes from 
$(v, v_n)=(194.27\mbox{GeV}, 173.75\mbox{GeV})$ to $(165.97\mbox{GeV}, 0)$ at $T_N$, and
from $(109.54, 0)$ to $(0,0)$ at $T_C$. The latter is a strongly first-order EWPT.\par
\begin{figure}[h!]
\centerline{\epsfig{file=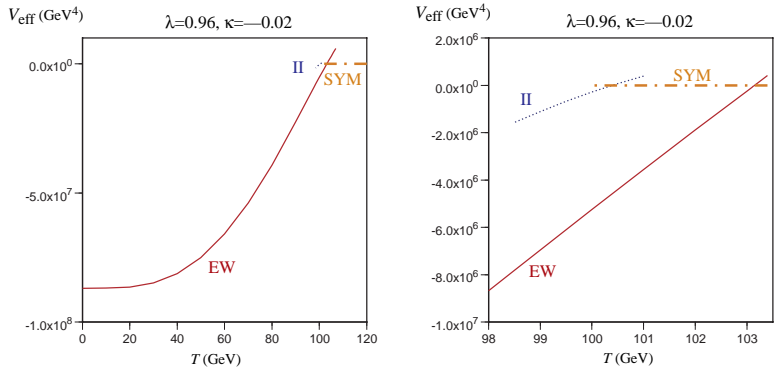, height=50mm}}
\caption{The effective potentials at the local minima corresponding to the phase-EW (solid curve),
the phase-II (dotted curve) and the phase-SYM (thick dotted-dashed curve) for the parameter set~D. 
The right-hand plot is the close view of the left-hand one near the transition temperature.}
\label{fig:Veffs-D}
\end{figure}

For the parameter set~D, the phase transition proceeds as shown in Fig.~\ref{fig:Veffs-D}.
Only the EWPT occurs at $T_C=103.14\mbox{GeV}$ where the order parameter changes
from $(v,v_n)=(182.49\mbox{GeV}, 192.26\mbox{GeV})$ to $(0,0)$.
This is a strongly first-order transition first studied in \cite{pietroni}.
Although there is also the local minimum corresponding to the phase-II in this case, it does
not take part in the phase transition.\par
The three parameter sets B, C and D are within the allowed region with light Higgs bosons. 
What discriminates the types of the transitions are the appearance of
the phase-\Iprime{}   and the relative magnitudes of the effective potential at the minima.
These factors depend on the parameters in the model.
All types of the transitions except for the type-B requires a light stop to admit the strongly first-order
EWPT. This is because the term which behaves as $T v^3$ with a negative coefficient is effective to make
the transition stronger for the three types, while the transition of type-B requires a local minimum
along $v=0$ and proceeds by leaping between the minimum and that corresponding to the phase-EW
so that the $v^3$-term is not necessary.
\subsection{CP-violating case}
The phases of the Higgs fields around the phase boundary are important ingredients which
determine how much baryon number is generated at the first-order EWPT\cite{reviewEWB}.
We found that the explicit CP violation in the squark sector weakens the $v^3$-behavior
of the effective potential\cite{FTT-1}, and expect that the same applies to the transitions which require
a light stop.
Here we study the effect of CP violation in the tree-level Higgs sector characterized by $\II$.
There are infinite sets of CP violating parameters which yield the same value of $\II$, so
we constrain them in such a way that the phase relevant to the neutron EDM, 
${\rm Arg}\lambda+\th_0+\varphi_0$, vanish\footnote{The EDM is an odd function of this phase
plus the phase of the $A$-term or the gaugino mass.}\cite{FunakuboTao}.
In practice, we set $\th_0=\varphi_0=0$ and take ${\rm Arg}\k$ as an independent 
parameter, from which ${\rm Arg}A_\k$ is determined by the conditions (\ref{eq:img-tadpole-1})
and (\ref{eq:img-tadpole-2}).\par
\begin{figure}[h!]
\centerline{\epsfig{file=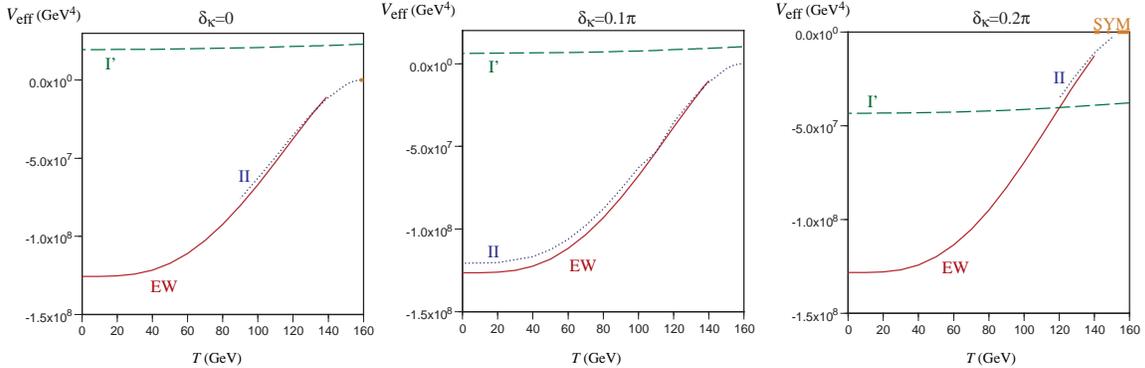, height=50mm}}
\caption{The effective potentials at the local minima corresponding to the phase-EW (solid curve),
the phase-\Iprime{}  (dashed curve), the phase-II (dotted curve) and the phase-SYM (thick dotted-dashed
curve) for $(\lambda,\k)=(0.83,-0.07)$ and $\delta_\k=0$, $0.1\pi$ and $0.2\pi$.}
\label{fig:Veffs-delk}
\end{figure}

Here we concentrate on the effect of the CP violation on the transition for the parameter sets
near the set~B, for which the EWPT is strongly first order even with heavy stops.
For the parameter set~B, we introduce the relative phase of $\k$ to $\k=-0.1$, $\delta_\k$, and 
repeat the numerical search for the minima at each temperature.
We find that the effective potential at the minimum corresponding to the phase-\Iprime{}
decreases with $\delta_\k$ and becomes smaller than that in the phase-EW even at zero temperature
for $\delta_\k\gtsim 0.2\pi$.
To see this behavior, we adopt another parameter set for which $\lambda=0.83$ and $\k=-0.07$ while
the others are the same as the set~B, and study temperature dependence of the effective potentail 
at the local minima. 
The masses and the couplings to the $Z$ boson of the Higgs mass eigenstates are listed in 
Table~\ref{tab:mass-g-delk}. Since $\absv\k$ is not so large, the effect of $\delta_\k$ on the mixing
of the CP-eigenstates is also small.
\begin{table}[h!]
\begin{center}
\begin{tabular}{c|c|ccccc}
\hline
   $\d_\k$  & & $H_1$ & $H_2$ & $H_3$ & $H_4$ & $H_5$    \\
\hline\hline
  \lowent{$0$} 
  & $m_{H_i}$(GeV) & $38.89$ & $75.31$ & $131.11$ & $625.61$ & $627.945$  \\
  & $g_{H_iZZ}^2$  & $6.213\times10^{-8}$  & $0$ & $0.9999$ & $6.816\times10^{-5}$ & $0$ \\
\hline
  \lowent{$0.1\pi$} 
  & $m_{H_i}$(GeV) & $40.04$ & $73.24$ & $131.20$ & $625.54$ & $627.56$  \\
  & $g_{H_iZZ}^2$  & $2.749\times10^{-6}$  & $0.00169$ & $0.9982$ & $6.570\times10^{-5}$ & $2.363\times10^{-6}$ \\
\hline
  \lowent{$0.2\pi$} 
  & $m_{H_i}$(GeV) & $43.21$ & $66.95$ & $131.38$ & $625.40$ & $627.85$  \\
  & $g_{H_iZZ}^2$  & $3.133\times10^{-5}$  & $0.00531$ & $0.9946$ & $6.132\times10^{-5}$ & $6.407\times10^{-6}$ \\
\hline
\end{tabular}
\end{center}
\caption{The mass and $g_{HZZ}^2$ of the Higgs mass eigenstates for $(\lambda,\k)=(0.83,-0.07)$
and $\d_\k=0$, $0.1\pi$ and $0.2\pi$.}
\label{tab:mass-g-delk}
\end{table}
The behaviors of the effective potential are shown in Fig.~\ref{fig:Veffs-delk}.
For $\delta_\k\gtsim 0.3\pi$, the zero-temperature vacuum is in the phase-\Iprime.   
The transitions for $\delta_\k=0$ and $0.1\pi$ are of type~C, while that for $\delta_\k=0.2\pi$ is
of type~B.
At the transition temperatures, the order parameters changes as follows:
\begin{tabbing}
 \hspace*{20mm} \= \hspace*{38mm} \= \hspace*{90mm} \kill
 $\delta_\k=0$;  
    \> at $T_N=133.22$GeV, \> $(v, v_n)=(180.74\mbox{GeV}, 195.49\mbox{GeV}) \rightarrow (163.16\mbox{GeV}, 0)$ \\
    \> at $T_C=158.27$GeV, \> $(v, v_n)=(26.33\mbox{GeV}, 0) \rightarrow (0, 0)$ \\[6pt]
 $\delta_\k=0.1\pi$;
    \> at $T_N=136.58$GeV, \> $(v, v_n)=(173.99\mbox{GeV}, 195.88\mbox{GeV}) \rightarrow (154.78\mbox{GeV}, 0)$ \\
    \> at $T_C=158.27$GeV, \> $(v, v_n)=(26.33\mbox{GeV}, 0) \rightarrow (0, 0)$ \\[6pt]
 $\delta_\k=0.2\pi$;
    \> at $T_C=120.16$GeV, \> $(v, v_n)=(200.62\mbox{GeV}, 208.93\mbox{GeV}) \rightarrow (0, 750.93\mbox{GeV})$
\end{tabbing}
The EWPT of the type~B for $\delta_\k=0.2\pi$ is strongly first order as that for the parameter set~B.
This behavior of the potential at the minimum in the phase-\Iprime{}   can be understood as follows.
As discussed in the previous section, the potential in this phase can be approximated by the tree-level
potential, which is given by (\ref{eq:hatV0-alpha}), even at finite temperatures.
In that expession, both $\hat R_\k$ and $m_N^2$ are decreasing functions of $\delta_\k\le\pi/2$, and
$\hat V_0(\alpha_+)$ decreases with $\delta_\k$\footnote{Note that $m_N^2$ is given by (\ref{eq:tadpole-mN}),
while we fixed $\absv\k$, $\absv\lambda$ and the charged Higgs mass, from which $R_\lambda$ is
determined.}.\par
The phase difference relevant to the baryogenesis is that at the phase transition after which the sphaleron
process decouples. 
We have considered such a transition to be the EWPT at which $v(T)$ acquires a nonzero value.
Even if the EWPT is strong enough to suppress the sphaleron process after that, no baryon number
can be generated at $T_C$ for the transition of type~C.
As we discussed in the previous section, there is the global $U(1)$ symmetry in the phase-II, so
that the phase of $v_2(T)+iv_3(T)$ is undetermined even in the presence of the explicit CP violation.
In fact, we observed many degenerate local minima with the same $\absv{v_2(T)+iv_3(T)}$ but
with different phases in the minimal search.
This implies that the bubble with different phases of the Higgs doublets are randomly created
at the EWPT. No net baryon number is generated, if such bubbles of a macroscopic number
are created as suggested by a naive argument of the transition\cite{carrington}.
If $v(T)$ at $T_C$ is too small to suppress the sphaleron process, 
just as in the case of $\delta_\k=0$ and $0.1\pi$ above, one may think that the second transition 
at $T_N$ will be another chance for the baryogenesis.
The same argument, however, applies to this case, so that the relative phase of the doublet,
as well as that of the singlet, between the two phases are undetermined. 
Thus the transition of type~C is not adequate for the baryogenesis.
On the other hand, the phase of the singlet is determined in the phase-\Iprime{}  in the transitions of type~B,
since the symmetry under phase rotation of the singlet field in the subspace of $v_1=v_2=v_3=0$ is
explicitly broken by the $R_\k$-term in the tree-level potential. 
In the example of the transition for $\delta_\k=0.2\pi$ above,
the $\varphi(T)$ changes from $0.0492$ (phase-EW) to $0.2154$ (phase-\Iprime).
Further the phase $\varphi(T)$  in the phase-\Iprime{}   is independent of temperature, since
all the field-dependent masses are independent of $\varphi=\tan^{-1}(v_5/v_4)$ in this phase,  so that
no $\varphi$-dependent terms are generated at finite temperatures.
\section{Discussions}
We studied various phase transitions in the NMSSM, characterized by the order parameters
$v(T)$ and $v_n(T)$, for the parameters consistent with the lower bound on the Higgs boson,
when the expectation value of the singlet scalar takes the value of the weak scale.
The phase transitions are classified into four types, one of which is the MSSM-like one, and
the others are peculiar to the NMSSM. These novel three types of transitions are realized
within the allowed region with a light Higgs boson. One of them is the transition studied
in \cite{pietroni}, but the other two types are equally possible.
We found that the transition of type~B can be strongly first order without need of a light stop.
For the parameter sets with this type of phase transitions, the Higgs bosons which is expected to 
be observed in future experiments is heavier than the present mass bound and has almost 
the same coupling to the $Z$ boson as the Higgs boson in the standard model.
Although the two light Higgs boson have very small couplings to the $Z$ boson,
their Yukawa couplings to the $b$ quark are the same order as the standard model Higgs boson,
so that they may be observed in hadron collider experiments.
In order to obtain the sphaleron decoupling condition, one must know the sphaleron solution
and its energy for the boundary conditions corresponding to the relevant phase transition.
A work in this direction is in progress and the result will be published elsewhere\cite{NMSSMsphal}.
The results suggest that the sphelaron does exist and its energy is a bit smaller than 
the original sphaleron in the standard model\cite{KM}, because of the negative cubic terms
in the potential.\par
We also invetigated the effect of the tree-level CP violation on the phase transitions of type~B
and C. Because of the global symmetry in the intermediate phase-II, the model which exhibits the transition
of type~C is not suited for the baryogenesis. On the other hand, the difference of $\varphi(T)$
above and below the transition temperature can be large without affecting the neutron EDM.
This phase becomes dependent on the spatial coordinate near the bubble wall created at the 
first-order transition. Noting that the $\mu$-parameter in the MSSM is effectively induced as
$\mu=\lambda v_n e^{i\varphi}/\sqrt2$, this introduces a new source of space-dependent phase
into the mass matrix of the charginos, neutralinos, squarks and sleptons, which will produce
new contributions to the source of the baryon number in the supersymmetric standard 
models\cite{EWBsusy}.\par
In this work, we have been focused on the static properties of the phase transitions.
To evaluate the generated baryon number, we must know about their dynamics.
Although the study of the dynamical aspects of the transitions is beyond the scope
of this work, we expect that as long as all the mass parameters in the Higgs potential are of the
order of the weak scale, the dynamics is not so far from those in the standard model or
the MSSM. In fact, we have not encountered an extremely deep minimum or high barriar 
between the minima of the effective potential in the numerical studies, which may
delay the transition or eternally trap the model within an unphysical state.
This situation might be altered, if we consider the case of larger $v_{0n}$ but not so large
as in the MSSM limit. It will be interesting to investigate the phase diagram of the model
for broader range of the parameters.
\section*{Acknowledgements}
The authors gratefully thank  A.~Kakuto and S.~Otsuki for valuable discussions.
This work was supported by the kakenhi of the MEXT, Japan, No.~13135222.
\appendix
\section{Notations}
\subsection{field-dependent masses}
Here we summarize the field-dependent masses $\barm$ used to define the effective potential
in terms of the vector $\bm{v}$ defined by (\ref{eq:def-vector-v}).
Those of the standard model particles are written by the components in the doublets as
\begin{eqnarray}
 \barm_Z^2 \!\!\!&=&\!\!\! {{g_2^2+g_1^2}\over4}(v_1^2+v_2^2+v_3^2),      \qquad
 \barm_W^2 = {{g_2^2}\over4}(v_1^2+v_2^2+v_3^2),        \\
 \barm_t^2 \!\!\!&=&\!\!\! {{\absv{y_t}^2}\over2}(v_2^2+v_3^2),     \qquad\qquad\quad
 \barm_b^2 = {{\absv{y_b}^2}\over2}v_1^2.
\end{eqnarray}
The those of the squarks and the singlet fermion depends on the singlet component $(v_4,v_5)$
and are given by
\begin{eqnarray}
 \barm_{\st_{1,2}}^2 \!\!\!&=&\!\!\!  
 {{m_\sq^2+m_{\st_R}^2}\over2}+{{g_2^2+g_1^2}\over{16}}(v_1^2-v_2^2-v_3^2)+{{\absv{y_t}^2}\over2}(v_2^2+v_3^2) \nonumber\\
 &&\!\!\!\!\!\!
 \pm\half\Biggl[
  \left(m_\sq^2-m_{\st_R}^2+{{x_t}\over2}(v_1^2-v_2^2-v_3^2)\right)^2           \\
 &&\!\!\!\!\!\!\!\!\!
  +2\absv{y_t}^2\left( {{\absv{\lam}^2}\over2}v_1^2(v_4^2+v_5^2)+\absv{A_t}^2(v_2^2+v_3^2)
                              -2\left[R_t(v_2v_4-v_3v_5)-I_t(v_3v_4+v_2v_5)\right]v_1 \right) \Biggr]^{1/2},     \nonumber\\
 \barm_{\sbo_{1,2}}^2 \!\!\!&=&\!\!\! 
 {{m_\sq^2+m_{\sbo_R}^2}\over2}-{{g_2^2+g_1^2}\over{16}}(v_1^2-v_2^2-v_3^2)+{{\absv{y_b}^2}\over2}v_1^2 \nonumber\\
 &&\!\!\!\!\!\!
 \pm\half\Biggl[
  \left(m_\sq^2-m_{\st_R}^2+{{x_b}\over2}(v_1^2-v_2^2-v_3^2)\right)^2           \\
 &&\!\!\!\!\!\!\!\!\!
  +2\absv{y_b}^2\left( {{\absv{\lam}^2}\over2}(v_2^2+v_3^2)(v_4^2+v_5^2)+\absv{A_b}^2 v_1^2
                              -2\left[R_b(v_2v_4-v_3v_5)-I_b(v_3v_4+v_2v_5)\right]v_1 \right) \Biggr]^{1/2},     \nonumber\\
 \barm_{\psi_N}^2 \!\!\!&=&\!\!\!  2\absv\k^2 (v_4^2+v_5^2),
\end{eqnarray}
where
\begin{eqnarray}
 x_t \!\!\!&=&\!\!\! {1\over4}\left(g_2^2-{5\over3}g_1^2\right),  \qquad
 x_b = -{1\over4}\left(g_2^2-{1\over3}g_1^2\right),      \\
 R_q \!\!\!&=&\!\!\! {1\over{\sqrt2}}{\rm Re}\left(\lam A_q e^{i(\th_0+\varphi_0)}\right), \qquad
 I_q =  {1\over{\sqrt2}}{\rm Im}\left(\lam A_q e^{i(\th_0+\varphi_0)}\right),  \qquad (q=t,b).
\end{eqnarray}
As for the convention for the squark sector, refer to \cite{FTT-1}.
\subsection{one-loop tadpole conditions}
The mass parameters in the tree-level potential is written in terms of the others by requiring that
the first derivatives of the zero-temperature effective potential evaluated at the vacuum be vanish.
We call these conditions as the tadpole conditions. The three conditions derived from the derivetives
with respect to the CP-even fields are used to express $m_1^2$, $m_2^2$ and $m_N^2$ as
\begin{eqnarray}
 m_1^2 \!\!\!&=&\!\!\! 
 \left(R_\lam-\half\RR v_{0n}\right) v_{0n}\tan\b_0-\half m_Z^2\cos(2\b_0) - {{\absv{\lam}^2}\over2}(v_{0n}^2+v_{0u}^2) \nonumber\\
 &&
 -{{N_C}\over{16\pi^2}}\Biggl[ 
   {{g_2^2+g_1^2}\over8} f_+\left(m_{\st_1}^2, m_{\st_2}^2\right) +
   {{t_1}\over{2v_{0d}\Delta m_\st^2}}f_-\left(m_{\st_1}^2, m_{\st_2}^2\right)
  + \left(\absv{y_b}^2-{{g_2^2+g_1^2}\over8}\right) f_+\left(m_{\sbo_1}^2, m_{\sbo_2}^2\right)          \nonumber\\
 &&\qquad\qquad\qquad
  + {{b_1}\over{2v_{0d}\Delta m_\sbo^2}}f_-\left(m_{\sbo_1}^2, m_{\sbo_2}^2\right) 
  - 2\absv{y_b}^2 m_b^2\left(\log{{m_b^2}\over{M^2}} -1 \right) \Biggr]       \nonumber\\
 &&
 -{3\over{32\pi^2}}\left[ {{g_2^2+g_1^2}\over2} m_Z^2\left(\log{{m_Z^2}\over{M^2}} -1 \right) +
                                  2\cdot{{g_2^2}\over2} m_W^2\left(\log{{m_W^2}\over{M^2}} -1 \right) \right],   \\
 m_2^2 \!\!\!&=&\!\!\! 
 \left(R_\lam-\half\RR v_{0n}\right) v_{0n}\cot\b_0+\half m_Z^2\cos(2\b_0) - {{\absv{\lam}^2}\over2}(v_{0n}^2+v_{0d}^2) \nonumber\\
 &&
 -{{N_C}\over{16\pi^2}}\Biggl[ 
  \left(\absv{y_t}^2-{{g_2^2+g_1^2}\over8}\right) f_+\left(m_{\st_1}^2, m_{\st_2}^2\right)
   + {{t_2}\over{2v_{0u}\Delta m_\st^2}}f_-\left(m_{\st_1}^2, m_{\st_2}^2\right)   \nonumber\\
 &&\qquad\qquad
   + {{g_2^2+g_1^2}\over8} f_+\left(m_{\sbo_1}^2, m_{\sbo_2}^2\right) 
   + {{b_2}\over{2v_{0u}\Delta m_\sbo^2}}f_-\left(m_{\sbo_1}^2, m_{\sbo_2}^2\right)  
   - 2\absv{y_t}^2 m_b^2\left(\log{{m_t^2}\over{M^2}} -1 \right) \Biggr]       \nonumber\\
 &&
 -{3\over{32\pi^2}}\left[ {{g_2^2+g_1^2}\over2} m_Z^2\left(\log{{m_Z^2}\over{M^2}} -1 \right) +
                                  2\cdot{{g_2^2}\over2} m_W^2\left(\log{{m_W^2}\over{M^2}} -1 \right) \right],   \\
 m_N^2 \!\!\!&=&\!\!\!
 \left( R_\lam - \RR v_{0n}\right){{v_{0d}v_{0u}}\over{v_{0n}}} + R_\k v_{0n} - {{\absv{\lam}^2}\over2}(v_{0d}^2+v_{0u}^2)
 -\absv{\k}^2 v_{0n}^2       \nonumber\\
 &&
 -{{N_C}\over{16\pi^2 v_{0n}}}\left[ 
   {{t_3}\over{2\Delta m_\st^2}} f_-\left(m_{\st_1}^2, m_{\st_2}^2\right) + 
   {{b_3}\over{2\Delta m_\sbo^2}} f_-\left(m_{\sbo_1}^2, m_{\sbo_2}^2\right)  \right] \nonumber\\
 &&
 + {{\absv\k^2}\over{4\pi^2}} m_{\psi_N}^2 \left( \log{{m_{\psi_N}^2}\over{M^2}} - 1 \right).     \label{eq:tadpole-mN}\\
\end{eqnarray}
The first derivatives with respect to the CP-odd fields lead to the relation among the CP-violating parameters
as
\begin{eqnarray}
 I_\lam \!\!\!&=&\!\!\!
 \half\II v_{0n}
 -{{N_C}\over{16\pi^2}}\left[ {{\absv{y_t}^2 I_t}\over{\Delta m_\st^2}} f_-\left(m_{\st_1}^2, m_{\st_2}^2\right)
                                      +{{\absv{y_b}^2 I_b}\over{\Delta m_\sbo^2}} f_-\left(m_{\sbo_1}^2, m_{\sbo_2}^2\right) \right],     
                                              \label{eq:img-tadpole-1}\\
 I_\k \!\!\!&=&\!\!\! -{3\over2}\II {{v_{0d}v_{0u}}\over{v_{0n}}}.    \label{eq:img-tadpole-2}
\end{eqnarray}
Here $t_i$ and $b_i$ are the components of the vector $\bm{t}$ and $\bm{b}$ defined in \cite{FunakuboTao},
respectively, and
\begin{equation}
 f_\pm(m_1^2,m_2^2) =  
 m_1^2\left(\log{{m_1^2}\over{M^2}} - 1\right) \pm m_2^2\left(\log{{m_2^2}\over{M^2}} - 1\right).
\end{equation}

%
%
%
%
%

%

\baselineskip=13pt
\begin{thebibliography}{99}
\bibitem{reviewEWB} For a review see, A.~Cohen, D.~Kaplan and A.~Nelson,
Ann. Rev. Nucl. Part. Sci. {\bf 43} (1993) 27.  \\
K.~Funakubo,\PrTP{96}{96}{475}.\\
V.~A.~Rubakov and M.~E.~Shaposhnikov, Phys. Usp. {\bf 39} (1996) 461 (hep-ph/9603208).
\bibitem{FunakuboTao} K.~Funakubo and S.~Tao, hep-ph/0409294.
\bibitem{carena} M.~Carena, J.~Ellis, A.~Pilaftsis and  C.~E.~M.~Wagner, Nucl. Phys. {\bf B586}
(2000) 92.
\bibitem{FTT-1} K.~Funakubo, S.~Tao and F.~Toyoda, Prog. Theor. Phys. {\bf 109} (2003) 415.
\bibitem{lightstop}   A.~Brignole, J.~R.~Espinosa, M.~Quiros and F.~Zwirner, Phys. Lett. \textbf{B324} (1994) 181.\\
  M.~Carena, M.~Quiros and C.~E.~M.~Wagner, Phys. Lett. \textbf{B380} (1996) 81.\\
  D.~Delepine, J.~M.~Gerard, R.~G.~Filipe and J.~Weyers, Phys. Lett. \textbf{B386} (1996) 183.\\
  J.~M.~Cline and G.~D.~Moore, Phys. Rev. Lett. \textbf{81} (1998) 3315.
\bibitem{pietroni} M.~Pietroni, Nucl. Phys. {\bf B402} (1993) 27.
\bibitem{davies} A.~T.~Davies, C.~D.~Froggatt and R.~G~.Moorhouse, Phys. Lett. {\bf 372} (1996) 88.
\bibitem{mssmEWPT} K.~Funakubo, Prog. Theor. Phys. {\bf 101} (1999) 415.\\
 M. Laine and K. Rummukainen, Nucl. Phys. {\bf B597} (2001) 23.
\bibitem{carrington} M.~E.~Carrington and J.~I.~Kapsta, Phys. Rev. {\bf D47} (1993) 5304.
\bibitem{NMSSMsphal} K.~Funakubo, A.~Kakuto, S.~Otsuki, S.~Tao and F.~Toyoda, in preparation.
\bibitem{KM} F.~R.~Klinkhammer and N.~S.~Manton, Phys. Rev. {\bf D30}
(1984) 2212.
\bibitem{EWBsusy} A.~G.~Cohen and A.~E~.Nelson, \PLB{297}{92}{111}.\\
P.~Huet and A.~E.~Nelson, \PRD{53}{96}{4578}.\\
M.~Carena, M.~Quir\'os, A.~Riotto, I.~Vilja and C.~E.~M.~Wagner,
\NPB{503}{97}{387}.\\
M.~P.~Worah, \PRLet{79}{97}{3810}.\\
M.~Aoki, N~.Oshimo and A.~Sugamoto, \PrTP{98}{97}{1179}.
%
\end{thebibliography}
\end{document}